\documentclass{article}
\usepackage{amsfonts}
\usepackage{amssymb}
\usepackage{amsmath}

\input{tcilatex}

\begin{document}

{\Large Some remarks about mass and Zitterbewegung }

\begin{center}
{\Large substructure of electron\medskip }
\end{center}

P. Brovetto$^{\dag }$, V. Maxia$^{\dag }$ and M. Salis$^{\dag \ddag }$ (%
\footnote{{\large ) E-mail: masalis@unica.it}})

$^{\dag }$On leave from Istituto Nazionale di Fisica della Materia -
Cagliari, Italy

$^{\ddag }$Dipartimento di Fisica - Universit\`{a} di Cagliari, Italy\medskip

\textbf{Abstrac}t - It is shown that the electron Zitterbewegung, that is,
the high-frequency microscopic oscillatory motion of electron about its
centre of mass, originates a spatial distribution of charge. This allows the
point-like electron behave like a particle of definite size whose
self-energy, that is, energy of its electromagnetic field, owns a finite
value. This has implications for the electron mass, which, in principle,
might be derived from Zitterbewegung physics.

Key words - Elementary particle masses, Zitterbewegung theory.\medskip

{\large Introduction}

The nature of mass of elementary particles is one of the most intriguing
topics in modern physics. Electron deserves a special attention because,
differently from quarks which are bound in composite objects (adrons), it is
a free particle. Moreover, differently from neutrinos whose vanishingly
small mass is still controversial, its mass is a well-known quantity.

At the beginning of the past century, the question of the nature of electron
mass was debated on the ground of the electromagnetic theory by various
authors (Abraham, Lorentz, Poincar\'{e}). Actually, the electron charge $e$
originates in the surrounding space an electromagnetic field $%
\overrightarrow{E},$ $\overrightarrow{H}$,\ associated with energy density $%
\left( E^{2}+H^{2}\right) /8\pi $ and momentum density $\left( 
\overrightarrow{E}\times \overrightarrow{H}\right) /4\pi c.$ These
densities, when integrated over the space, account for electron self-energy
and inertial mass. However, peculiar models must be introduced in order to
avoid divergent results. The simplest assumes that charge $e$ is spread on
the surface of a sphere of radius $r_{0}$. Consequently, an electron at rest
originates the electric field $\overrightarrow{E}=e\overrightarrow{r}/r^{3}$
for $r\geq r_{0}$ and $\overrightarrow{E}=0$ for $r<r_{0}$. So, electron
self-energy is found to be%
\begin{equation}
w_{e}=\frac{1}{2}\dint\limits_{r_{0}}^{\infty }\frac{e^{2}}{r^{2}}dr=\frac{%
e^{2}}{2r_{0}}.  \label{pat}
\end{equation}%
By putting: $w_{e}=m_{e}c^{2}$ ($m_{e}$ standing for the electron rest mass: 
$9.1\cdot 10^{-28}$ g) and $2r_{0}=r_{e}$, we get:$\
r_{e}=e^{2}/m_{e}c^{2}=2.8\cdot 10^{-13}$ cm. This quantity is referred to
as the classical electron radius $\left[ 1\right] $.

The advent of quantum physics made obsolete these models. In quantum
electrodynamics, electrons are considered structureless point-like
particles. Therefore, in order to\ evade the\ consequent divergence\
problem, renormalization methods were introduced which regard the divergent
electron mass as an unobservable quantity. On that account, empirical mass
is utilized in applications.\ Nowadays, keeping this state of affairs in
mind, most theoretical physicists ascribe particle masses to interactions
with the Higgs field, that is, a hypothetical field associated with
zero-spin bosons, which permeates the universe. Accordingly, electron mass
is assumed proportional to the Higgs vacuum expectation value $v,$ that is, 
\begin{equation}
m_{e}=g_{e}v/\sqrt{2},  \label{cal}
\end{equation}%
$g_{e}$ standing for a coupling constant specific for the electron $\left[ 2%
\right] $. This picture provides a physical basis for particle masses, but,
owing to the unknown values of coupling constants, actual masses remain
empirical.\medskip

{\large The electron Zitterbewegung}

Electron is rightly described by the relativistic Dirac equation. Besides
the spin moment\ $\overrightarrow{S}$,\ which is represented by a rank-3
four-dimensional antisymmetric tensor, electron shows an electromagnetic
moment whose density-components are represented by an antisymmetric rank-2
four-dimensional antisymmetric tensor $\left[ 3\right] .$ In the reference
frame in which electron is at rest the three magnetic components, when
integrated over space, yield a moment $\overrightarrow{\mathfrak{M}}$
related to spin by

\begin{equation}
\overrightarrow{\mathfrak{M}}=-\frac{e}{m_{e}c}\overrightarrow{S},
\label{xa}
\end{equation}%
which, by putting $\mathfrak{M=}\left\vert \overrightarrow{\mathfrak{M}}%
\right\vert $, gets

\begin{equation}
\mathfrak{M=}eh/4\pi m_{e}c=\mathfrak{\mu }_{B},  \label{xb}
\end{equation}%
$\mathfrak{\mu }_{B}$ standing for the Bohr magneton. Performing \ a Lorentz
transformation, owing to the different tensorial characters, $%
\overrightarrow{\mathfrak{M}}$ and $\overrightarrow{S}$ are no longer
parallel and\ $\mathfrak{M}$ is only approximately equal to the Bohr magneton%
$.$ So, an electron, in its ground state moving in the field of a charge $Ze$%
, shows the magnetic moment

\begin{equation}
\mathfrak{M=\mu }_{B}\left( 1+2\sqrt{1-\alpha ^{2}Z^{2}}\right) /3,
\label{xc}
\end{equation}%
$\alpha $ standing for the fine structure constant $\left[ 4\right] .$ The
three electric components of moment tensor yield\ in turn a dipole moment $%
\overrightarrow{\mathcal{P}}$ related to\ magnetic moment by

\begin{equation}
\overrightarrow{\mathcal{P}}=\overrightarrow{\mathfrak{M}}\times \frac{%
\overrightarrow{v}}{c}  \label{cel}
\end{equation}%
$\left[ 5\right] $ (\footnote{%
) Really, the electron dipole moment was first derived by J.\ Frenkel \ on
the ground of mere relativistic considerations.}). These features of the
electromagnetic tensor\ hardly are compatible with a point-like electron.
They fit better with an electron showing an internal structure.

Still, another important property, which was named Zitterbevegung by Schr%
\"{o}- dinger (\footnote{%
) Zitterbewegung can be translated: "Jitterbehaviour".}), derives from Dirac
equation when dealing with expected electron velocity $\left[ 6\right] $.
Different treatments have been given on this subject $\left[ 7,8,9\right] $.
We consider here the treatment, given by De Broglie, which is especially
convenient for our purposes $\left[ 3\right] .$ In Dirac theory, operators
for $x$, $y$,\ $z$ velocity components are $-c\mathbf{\alpha }_{1},$ $-c%
\mathbf{\alpha }_{2},$ $-c\mathbf{\alpha }_{3\text{,}}$ respectively. So,
expected velocity on $x$ axis is%
\begin{equation}
\left\langle \overset{\cdot }{x}\left( t\right) \right\rangle =\diiint
\sum_{k=1}^{k=4}\psi _{k}^{\ast }\left( r,t\right) \left( -c\mathbf{\alpha }%
_{1}\right) \psi _{k}\left( r,t\right) d^{3}\overrightarrow{r},  \label{pet}
\end{equation}%
$\psi _{k}$ standing for the spinor components. These are given in general
by the Fourier's expansion%
\begin{equation}
\psi _{k}\left( r,t\right) =\diiint \left\{ a_{k}\left( \overrightarrow{p}%
\right) \exp \left[ \left( 2\pi i/h\right) \left( Wt-\overrightarrow{p}\cdot 
\overrightarrow{r}\right) \right] +\right.  \label{pit}
\end{equation}

\begin{equation*}
\left. +b_{k}\left( \overrightarrow{p}\right) \exp \left[ \left( 2\pi
i/h\right) \left( -Wt-\overrightarrow{p}\cdot \overrightarrow{r}\right) %
\right] \right\} d^{3}\overrightarrow{p},
\end{equation*}%
where $a_{k}$ and $b_{k}$ mean the amplitudes of positive and negative
energy states, respectively. The latter cannot be omitted owing to the
completeness theorem of Fourier's expansion. Fermi actually showed that
without the contribution of negative energy states a free electron cannot
originate Thomson's scattering of light. This scattering appears as a sort
of resonance of the quantum\ jump of energy $2m_{e}c^{2}$ between positive
and negative energy states $\left[ 10\right] $. Substituting eq. (\ref{pit})
into eq. (\ref{pet}), a sum of conjugated terms is found, representing an
oscillation of angular frequency $4\pi W/h$, that is,%
\begin{equation}
-c\sum_{k=1}^{k=4}\left\{ a_{k}^{\ast }\mathbf{\alpha }_{1}b_{k}\exp \left[
-\left( 4\pi i/h\right) Wt\right] +b_{k}^{\ast }\mathbf{\alpha }%
_{1}a_{k}\exp \left[ \left( 4\pi i/h\right) Wt\right] \right\} =  \label{bum}
\end{equation}

\begin{equation*}
=cA_{1}\cos \left( \frac{4\pi W}{h}t+\varphi _{1}\right) ,
\end{equation*}%
where\ amplitude $A_{1}$ and phase $\varphi _{1}$ depend on moment $%
\overrightarrow{p}.$ The electron Zitterbewegung consists just of this
oscillatory behaviour of velocity. It is due to the beat between positive
and negative energy states. This result entails that the Ehrenfest theorem
is not right in relativistic quantum mechanics. Performing integration with
respect to time, straightforward transformations lead to 
\begin{equation}
\left\langle x\left( t\right) \right\rangle =\left\langle v_{x}\right\rangle
t+\sum_{\sigma }\sigma \frac{hc}{4\pi W}A_{1}\sin \left( \frac{4\pi W}{h}%
t+\varphi _{1}\right) ,  \label{pot}
\end{equation}%
where 
\begin{equation}
\left\langle v_{x}\right\rangle =\sum_{\sigma }\sigma \frac{p_{x}c^{2}}{W}%
\sum_{k=1}^{k=4}\left( a_{k}^{\ast }a_{k}^{{}}-b_{k}^{\ast }b_{k}^{{}}\right)
\label{put}
\end{equation}%
is the time-independent part of $x$-component of velocity. In previous
equations $\sigma $ stands for the element $dp_{x}dp_{y}dp_{z}$ of momentum
space. Amplitude $A_{1}$ and phase $\varphi _{1}$, like coefficiens $a_{k}$
and $b_{k}$,\ vary from one element $\sigma $ to the other. Analogous
results are found for $y$ and $z$ components.

\bigskip

{\large The electron substructure}

We are dealing here with the electron rest mass. We assume, therefore, that
energy $W$ barely exceeds $m_{e}c^{2}.$ So, terms dependent on moment $%
\overrightarrow{p}$ can be disregarded in comparison with $m_{e}c$. By
putting \ $W=m_{e}c^{2}$ and omitting time-independent velocity components,
previous results allow us to write%
\begin{equation*}
\left\langle x\left( t\right) \right\rangle =\frac{\lambda _{e}}{4\pi }%
\sum_{\sigma }\sigma A_{1}\sin \left( \frac{4\pi }{T_{Z}}t+\varphi
_{1}\right) ,
\end{equation*}%
\begin{equation*}
\left\langle y\left( t\right) \right\rangle =\frac{\lambda _{e}}{4\pi }%
\sum_{\sigma }\sigma A_{2}\sin \left( \frac{4\pi }{T_{Z}}t+\varphi
_{2}\right) ,
\end{equation*}%
\begin{equation}
\left\langle z\left( t\right) \right\rangle =\frac{\lambda _{e}}{4\pi }%
\sum_{\sigma }\sigma A_{3}\sin \left( \frac{4\pi }{T_{Z}}t+\varphi
_{3}\right) ,  \label{cil}
\end{equation}%
where $\lambda _{e}=h/m_{e}c$ is the Compton wavelength and%
\begin{equation}
T_{Z}=\frac{h}{m_{e}c^{2}}=8.1\cdot 10^{-21}\text{s}  \label{zac}
\end{equation}%
is the time light requires to cover a space equal to Compton wavelength. It
is twice the Zitterbewegung period. According to eqs. (\ref{cil}), electron
motion consists of a superposition of oscillations of different amplitudes
and phases. If\ oscillations on $x$, $y$, $z$ axes have the same phase,
electron\ moves along a $\overrightarrow{s}$ axis passing through the
origin. If also amplitudes have a common value $A$, electron\ moves along
the $\overrightarrow{s}$ axis performing \ angle $54.7%
{{}^\circ}%
$ with respect to axes, that is,

\begin{equation}
s=\sqrt{3}\Lambda \sin \left( \frac{4\pi }{T_{Z}}t\right) .  \label{col}
\end{equation}%
where $\Lambda =\lambda _{e}\sigma A/4\pi $. Owing to factor $\sigma A/4\pi $%
, length\ $\Lambda $ is expected to be small in comparison with Compton
wavelength. If amplitudes are equal but different phases: $\varphi
_{1}=\varphi _{3}=\pi /2$, $\varphi _{2}=0$ are considered, electron
performs a turn in $x$, $y$ plane accompanied by an oscillation along $z$
axis, that is, utilizing cylindrical coordinates $r$, $\vartheta $, $z$,

\begin{equation}
r=\Lambda ,\qquad \vartheta =4\pi \frac{t}{T_{Z}},\qquad z=\Lambda \cos
\left( \frac{4\pi }{T_{Z}}t\right) .  \label{cul}
\end{equation}%
The electron moves on a trajectory wrapped round a cylinder of radius $%
\Lambda $ and height $2\Lambda $.

Previous equations show that the oscillating electron owns a dynamical
substructure. But, the uncertainty principle sets restrictions to electron
oscillations. In fact, by dividing\ relation $\delta w\cdot \delta t\simeq h$
by $m_{e}c^{2}$, we obtain%
\begin{equation}
\frac{\delta w}{m_{e}c^{2}}\cdot \frac{\delta t}{T_{Z}}\simeq 1.  \label{cac}
\end{equation}%
Owing to this reciprocity law, when electron energy $w$ is determined with a
precision such that\ $\delta w<<m_{e}c^{2}$, indetermination on time is $%
\delta t>>T_{Z}$. In this case the oscillating motion of electron cannot be
observed. This occurs, for instance, when dealing with atomic spectroscopy.
In fact, the most important quantity which controls precision in atomic
spectroscopy is the Rydberg constant$\ R_{H}=13.6056981\unit{eV}$. It is
known with seven digits, corresponding to an indetermination $\delta w$
smaller than $10^{-7}$ $\unit{eV}$. Taking into account that $m_{e}c^{2}=0.5$
MeV, we have $\delta w/m_{e}c^{2}\leq 2\cdot 10^{-13},$ so that $\delta t$/$%
T_{Z}\geq 5\cdot 10^{12}$. This large indetermination in time forces us to
eliminate\ $t$ in previous equations. Let consider, for instance, the
oscillation given in eq. (\ref{col}). In a half-period electron moves from $%
s=-\sqrt{3}\Lambda $ to $s=+\sqrt{3}\Lambda $. Therefore, probability $dP$
that electron is found in the space between $s$ and $s+ds$ is: $dP=4dt\
/T_{Z}$, $dt$ standing for the time required to cross space $ds$. We have
thus: $dP/ds=4\ /\left( \overset{\cdot }{s}T_{Z}\right) $. Calculating $%
\overset{\cdot }{s}$ and eliminating $t$, we get

\begin{equation}
\frac{dP}{ds}=\frac{1}{\pi \sqrt{3\Lambda ^{2}-s^{2}}}.  \label{cec}
\end{equation}%
Analogous considerations apply to trajectories (\ref{cul}). In this case, a
complete period is required to cover the trajectory, which leads to: $%
dP/ds=2\ /\left( \overset{\cdot }{s}T_{Z}\right) $. We have from eqs. (\ref%
{cul})

\begin{equation}
ds^{2}=\Lambda ^{2}d\vartheta ^{2}+dz^{2}=\Lambda ^{2}\left[ 1+\sin
^{2}\left( \frac{4\pi }{T_{Z}}\right) \right] dt^{2},  \label{cic}
\end{equation}%
which leads to

\begin{equation}
\frac{dP}{ds}=\frac{1}{2\pi \sqrt{2\Lambda ^{2}-z^{2}}}\text{.}  \label{coc}
\end{equation}%
The linear oscillation of eq. (\ref{cec}) differs from the cylindrical
oscillation of eq. (\ref{coc}) because the former allows charge to be found
at the axis origin, while the second places charge only on the cylinder
surface. Taking into account that in eqs. (\ref{cil}) a number of
oscillations are added up, this is tantamount to say that the oscillating
electron originates a spatial distribution of charge. Linear oscillations
originate a central distribution of charge, while cylindrical oscillations
on $x$, $y$, and $z$ axes originate an empty shell of charge. This is a very
interesting feature, because such shell mimics the distribution of charge on
the spherical surface postulated in eq. (\ref{pat}).

It is to be emphasized that the foregoing results are not in conflict with
the point-like nature of electron observed in high-energy experiments, such
as electron-positron collisions. Late experiments of this kind have shown
that electron radius does not exceed $10^{-16}$ cm, that is, at least three
orders of magnitude less than the classical radius. In these experiments\
the collision time $\tau $, as determined in the reference frame of the
electron, is given by the ratio between the impact parameter, that is, the
electron size\ $10^{-16}\ $cm, and the velocity of light $c$, that is, $\tau
=3\cdot 10^{-27}$ s. This time is small with respect to $T_{Z}$. In
addition,\ Lorentz contraction along the direction of collision makes
sharper the electric field pulse originated by the colliding positron so
further reducing $\tau $. It follows that eqs. (\ref{cec}, \ref{coc}) cannot
be applied.

We come now at the question of electron mass. The opinion that the
Zitterbewegung substructure has implications for electron self-energy, that
is, for electron\ rest mass, is not new. It was first advanced by A. O.
Barut and A. J. Bracken $\left[ 8\right] $. These workers, however, left out
the possibility that this dynamical substructure originates a spatial\
distribution of charge. It, in principle, allows evaluation of electron
field and consequently of its self-energy, like in eq. (\ref{pat}). For
clarity sake, let consider in succession the following steps.

$1%
{{}^\circ}%
)$ Amplitudes and phases in eqs. (\ref{cil}) are found on the ground of the
Zitterbewegung theory. Then, by proceeding as in eqs. (\ref{cec}, \ref{coc}%
), electron oscillations are transformed in a spatial distribution of charge.

$2%
{{}^\circ}%
)$ Evaluation of field and self-energy: $w_{e}=w_{Z}\left(
m_{e},h,c,e\right) $ of this distribution.

$3%
{{}^\circ}%
)$ After substitution of $m_{e}$ with a variable "trial mass" $\mu $,
electron self-energy is evaluated again as a function of $\mu $. Then,
function $w_{Z}\left( \mu ,h,c,e\right) $ is compared with trial energy $\mu
c^{2}$. If equation

\begin{equation}
w_{Z}\left( \mu ,h,c,e\right) =\mu c^{2}  \label{puz}
\end{equation}%
has a solution $\mu =\mu _{0}$ different from zero, it follows that $%
m_{e}=\mu _{0}$ is the expected electron mass given as a quantity dependent
on the fundamental constants $h$, $c$, $e$. It is worth to point out in this
connection that possible further solutions of eq. (\ref{puz}) might\ be
interpreted as the high-mass flavours, $mu$ and $tau$, of electron.

\bigskip

{\large Discussion and conclusions}

Unfortunately a thorough Zitterbewegung theory has not yet been expressed.\
Apart from amplitudes and phases, different points would require a more\
deep analysis. Perhaps, it might be necessary to interpret Dirac's equation
as a quantized field equation, rather than at the one-particle level $\left[
8\right] $.\ This poor state of affairs is due to the fact that most
theorists relegate Zitterbewegung to the position of an unobservable
mathematical curiosity. Consequently, evaluation of actual electron mass is
left out for now. Nevertheless, the previous qualitative arguments are
interesting because they show that in low energy experiments the point-like
electron takes a spatial substructure. Moreover, they show that it is
possible, in principle, to deal with the problem of particle masses without
introducing new physics as occurs with the Higgs-field theory. The last, on
the other hand, is somewhat disappointing because the assignment of the
coupling costants to particles is arbitrary. This omits explaining why
masses of particles are so different. On the contrary, according to
Zitterbewegung theory,\ particle masses come from\ interaction with their
own fields, so that neutrinos, which show only the short-range weak\ field,
are the less massive. Mass of electrons which show both weak and
electromagnetic interactions follows. Quarks, which combine all kinds of
interactions, are the most massive particles.

As for the range of weak forces, Higgs-field theory ascribes such very short
range to the high masses of weak bosons $W^{+}$,\ $W^{-}$ and $Z^{0}$. This
poses the problem of the nature of these masses. In the Higgs-field theory,
when the boson system is in its ground state, massless weak bosons acquire
mass through a mechanism referred to as spontaneous symmetry breaking.
Unlike this interpretation, we proposed in a previous paper that weak bosons
remain massless and the short range of weak interactions is due to the
screening properties of the negative-kinetic-energy distribution of
neutrinos that, according to Dirac's equation, permeates the universe $\left[
9\right] $. The general conclusion ensues: all field-bosons are massless and
all fermions acquire mass owing to the fields themselves originate.

\bigskip

{\large References}

$\left[ 1\right] $ See for instance: L. Page and N. I. Adams -
Electrodynamics (Dover, 1965) Ch. 4.

$\left[ 2\right] $ G. Kane - Modern Elementary Particle Physics
(Addison-Wesley, 1987) Ch. 8.

$\left[ 3\right] $ L. De Broglie - L'Electron Magnetique (Hermann, Paris,
1934) Ch. XXI.

$\left[ 4\right] $ G. Breit - Nature \textbf{122}, 649 (1928).

$\left[ 5\right] $ J. Frenkel - Zeits. f. Phys., \textbf{37}, 243 (1926).

$\left[ 6\right] $ E. Schr\"{o}dinger, Sitzungsb. Preuss. Akad. Wiss.
Phys.-Math. Kl. \textbf{24}, 418 (1930); \textbf{3}, 1 (1931).

$\left[ 7\right] $ P. A. M. Dirac - The Principles of Quantum Mechanics
(Oxford, 1958) \S\ 69.

$\left[ 8\right] $ A. O. Barut and A. J. Bracken, Phys. Rev. D \textbf{23,}
2454 (1981).

$\left[ 9\right] $ B. R. Holstein - Topics in Advanced Quantum Mechanics
(Addison-Wesley, 1992) Ch. VI.

$\left[ 10\right] $ E. Fermi - Rev. Mod. Phys. \textbf{4}, 120 (1932).

$\left[ 11\right] $ P. Brovetto, V. Maxia and M. Salis - Il Nuovo Cimento A 
\textbf{112}, 531 (1999).

\end{document}